\documentclass[twocolumn,letterpaper,fleqn]{ModDetSymp}
\usepackage{fix2col}
\usepackage{multicol}
\usepackage{ifthen}
\usepackage{amsmath}
\usepackage{graphicx}
\usepackage{xcolor}
\usepackage{calrsfs}

\title{The Salinas Reactive-Flow Rate Model}

\author{Peter~T.~Williams$^{\dag}$}

\affiliation{$^{\dag}$Mechanical Systems Department, The Aerospace Corporation, United States\\}

\begin{document}

\twocolumn[
\setlength{\fboxrule}{0.5pt}
\begin{@twocolumnfalse}
\maketitle
\begin{center}
\parbox{5in}{\textbf{Abstract. }%
We present the model Salinas and discuss its results in modeling corner-turning behavior, including failure and the formation of dead zones. Salinas is a recent reactive-flow model for detonation in insensitive high explosives, inspired by JWL++ models but informed by a range of other models suited for corner turning. The model is computationally efficient and has a minimum of free parameters. 

\sectionline}
\end{center}
\end{@twocolumnfalse}
]

\section{Introduction}
The detonation of insensitive high explosives (IHE) poses many challenges to modelers. Corner turning is one such problem. A detonation wave travelling around a corner through IHE (with or without confinement) may exhibit failure, leading to the formation of dead zones in which the IHE fails to detonate. Accurate, efficient and trustworthy prediction of this phenomenon is of great interest both to modelers and the consumers of those models alike, but has proved difficult. 

We present Salinas, a reactive-flow model for the modeling of detonation in IHE.  Salinas can successfully capture corner-turning including failure and the formation of dead zones. The model is based on JWL++ models, using a Jones-Wilkins-Lee (JWL) equation of state (EOS) for the fully reacted products and a simple two-parameter Murnaghan EOS for the unreacted high explosive. It was inspired by the JWL++ Tarantula model developed at Lawrence Livermore National Laboratory (LLNL), as well as by the Ignition \& Growth model, the Statistical Hot Spot model, and CREST. 


\section{Informal Development}
Let a passing strong shock initiate burning at a number density $\chi_*$ of active hotspots.
Assume that burning starts at inifinitesimal points and proceeds in spherical burn fronts that grow and overlap. Poisson statistics shows that the ratio of surface area to volume for the burning surface is
\begin{equation}
\frac{A}{V} = \left(36\pi\right)^{1/3} \chi_*^{1/3} \left[-\ln(1-F)\right]^{2/3}(1-F)
\label{eq:AtoVratio}
\end{equation}
where $F$ is the burn fraction, progressing $0\rightarrow 1$. 

(Note that $(36\pi)^{1/3}$ is the surface area of a sphere of unit volume, $\chi_*^{-1/3}$ is 
a lengthscale, and $\left[-\ln(1-F)\right]^{2/3}(1-F) \approx F^{2/3}$ when $F$ is small. As
$F$ gets larger, the statistics of the probability of overlapping spherical burn fronts becomes
progressively more significant.)

To develop a rate model for reactive flow, we only need to know three additional things: (A) what is the speed of the mesoscopic deflagration front, (B) how do we quantify a shock's ability to 
intiate burn at potential hotspots, (C) given a passing shock, how many actual active hotspots are there?

In fact, in answer to (A), we can even claim ignorance of the actual speed of deflagration, and only make use of how it {\em scales}. And the simplest answer here, in keeping with a very long line
of investigation of energetic material, is to assume that deflagration speed is a simple power law
with thermodynamic pressure, so that burn rate is proportional to $P^b$ for some $b$. To make it
even simpler, we settle on the commonly-adopted value $b=2$.

For (B), we measure the ``strength'' of a shock --- meaning, in an informal way, its ability to initiate burning at potential hotspots --- by the peak pressure attained in the shock. This is not perfect; we could imagine various objections
--- for example, we might ask how quickly the peak pressure is attained, and how long does the pressure stay near its peak --- valid questions, to be sure --- but, we argue, it is good enough.
We follow JWL++ Tarantula 2011 terminology and call this ``max historical $P$'' or $P_{\rm mh}$.

Finally, having settled on (B), we ask what is the distribution of sensitivity to potential hot spots, in terms of $P_{\rm mh}$? We need some PDF of the likelihood that a given shock of strength $P_{\rm mh}$ will initiate burn at a potential hotspot. That is, we need some probability measure for how easy or difficult it may be to initiate burn at each of our potential hotspots. Some potential hotspots will be quite sensitive; others, less so.

We argue that the most physically and mathematically natural assumption, given what we know about
the sensitivity of IHE ({\em e.g.} TATB) to initiation --- which seems repeatedly to invite modelers to use sigmoid functions like $\tanh$ --- is that the sensitivity of potential hotspots is lognormally distributed. In other words, the PDF of sensitivity, plotted in the logarithm of $P_{\rm mh}$, is Gaussian. This seems reasonable if we invoke the Central Limit Theorem and
assume that the sensitivity of any given potential hotspot is the product of a large number of more or less uncorrelated random factors.

We then have an overall rate law that may be written
\begin{equation}
\frac{DF}{Dt} = G \cdot {\cal R}(P_{\rm mh}) \cdot {\cal P}(P) \cdot {\cal F}(F).
\end{equation}
Individually, each of the component functions is as follows. First, ${\cal R}$ is just
a standard lognormal CDF raised to the $1/3$ power:
\begin{equation}
{\cal R}(P_{\rm mh}) = \Phi^{1/3}\left(\frac{\ln P_{\rm mh} - \ln P_\mu}{\sigma} \right)
\end{equation}
where
\begin{equation}
\Phi(x) = \frac{1}{2}\left[1+{\rm erf}\left(\frac{x}{\sqrt{2}} \right) \right],
\end{equation}
and $P_\mu$ and $\sigma$ are model parameters that characterize the HE's hotspot sensitivity.
The $1/3$ power just comes from its appearance
in the surface-to-volume ratio in eq.~(\ref{eq:AtoVratio}). This exponent may seem counterintuitive, but 
note that the rate is still first order in $\Phi$ with respect to {\em time}; the cube root comes in because we 
are writing the rate as a function of $F$, not $t$.

Given a passing shock of strength
$P_{\rm mh}$ and a number density of potential hotspots $\chi$, the number density of actual
active hotspots $\chi_*$ is just
\begin{equation}
\chi_* = \chi {\cal R}(P_{\rm mh}).
\end{equation}

\begin{figure}
\includegraphics[width=7.0cm]{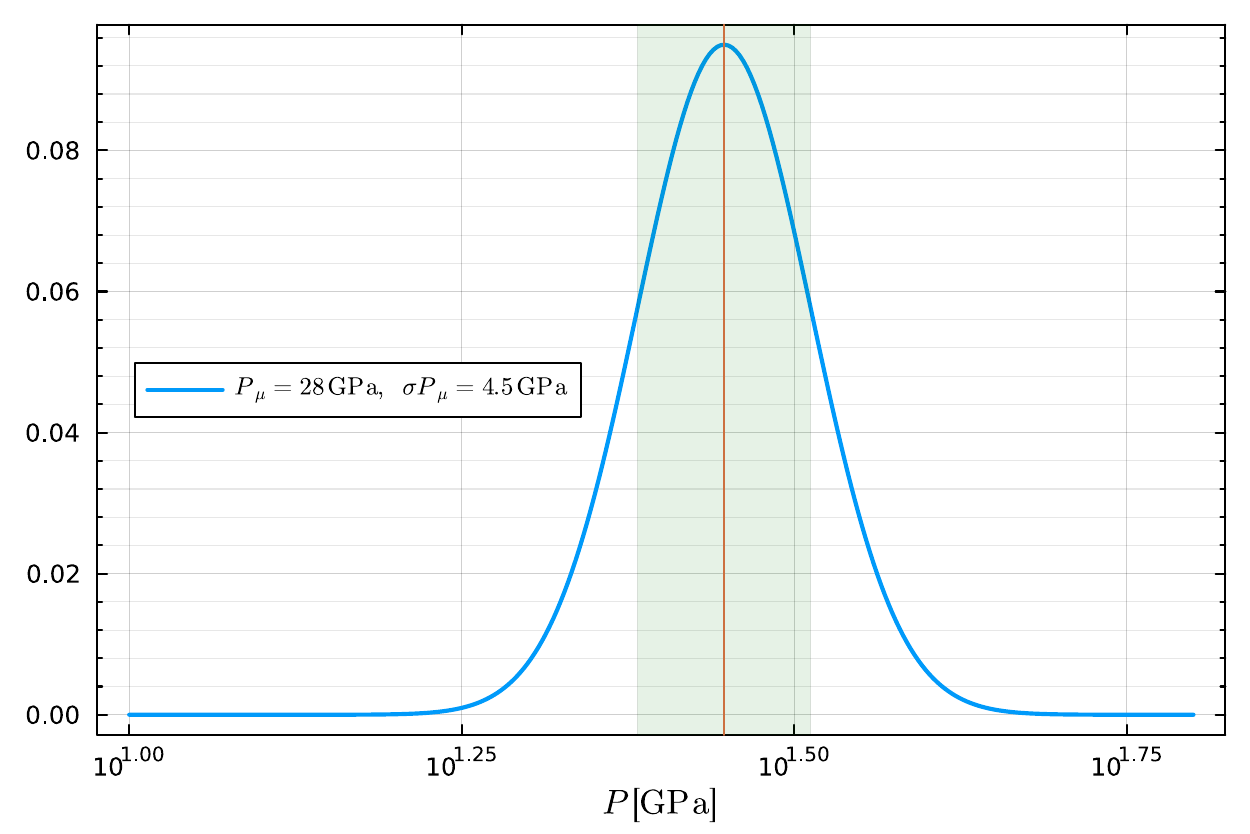}
\caption{Sample lognormal PDF of hot-spot sensitivity. Shaded area is $\pm 1\,\sigma = 0.15$.
Vertical red line is $P_\mu$.}
\label{fig:lognormal} 
\end{figure}

The pressure scaling function as described above is just
\begin{equation}
{\cal P}(P) = \left( \frac{P}{P_0}\right)^b
\end{equation}
where $P_0$ is just some reference pressure. In previous work \cite{Will_2020} we chose $P_0=P_{\rm CJ}$, which has 
advantages and disadvantages, since the CJ pressure is not necessarily well-known. This choice makes it possible to relate the mesoscopic deflagration speed $u$ to its value at the CJ point:
\begin{equation}
u = u_{\rm CJ} {\cal P}(P).
\end{equation}
This may prove valuable in trying to connect model parameters to mesoscopic quantities, but it is not
really necessary for the purpose of developing the model itself.

\begin{figure}
\includegraphics[width=7.0cm]{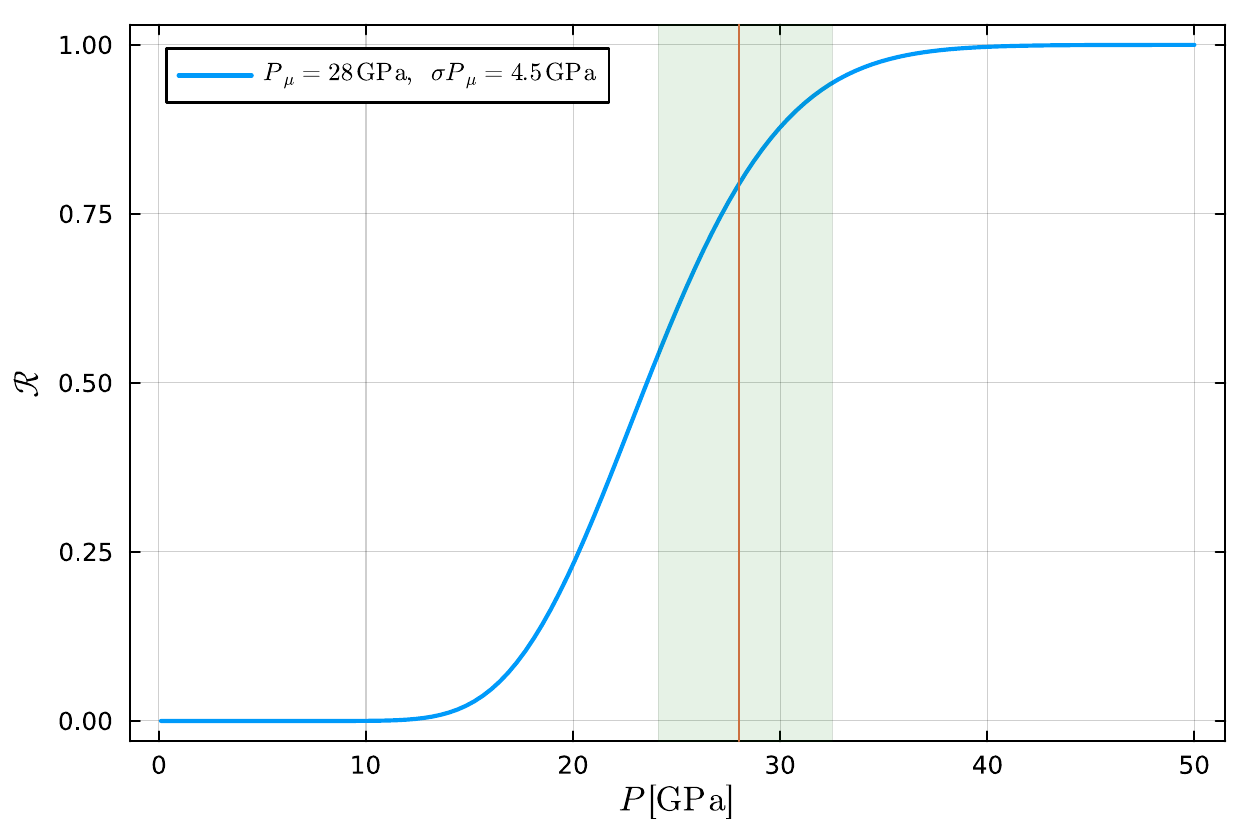}
\caption{Sample initiation function ${\cal R}(P_{\rm mh}$. Shaded area corresponds to
shaded area in fig.~(\ref{fig:lognormal}). Vertical red line is $P_\mu$.}
\label{fig:SalinasRfunc} 
\end{figure}

The form-factor function ${\cal F}$ is
\begin{equation}
{\cal F}(F) = \left[-\ln(1-F)\right]^{2/3}(1-F)
\end{equation}
and has no free parameters, being derived directly from geometry and statistics.

Finally, just as we can claim ignorance of the deflagration speed $u_{\rm CJ}$ at the CJ point, we do not need to know the actual number density $\chi$ of potential hotspots, but we can relate the overall
prefactor $G$ to the product $\chi^{1/3} u_{\rm CJ}$:
\begin{equation}
G = \left( 36 \pi \right)^{1/3} \chi^{1/3} u_{\rm CJ}.
\end{equation}

In actual practice, of course, one typically 
does not attempt to measure $\chi$ or $u_{\rm CJ}$, but simply
adopts a value for $G$ in the course of fitting simulation to experiment.
Typically, for PBX~9502, we found values of $G$ on the order of 
roughly $50\,{\rm \mu s}^{-1}$ to work well.

In some instances it may prove
more convenient to fold the reference pressure $P_0$ ({\em e.g.} $P_{\rm CJ}$) into the rate
constant $G$. Let us put primes on $G$ and ${\cal P}$ to indicate this alternative way of writing
the rate, in which ${\cal P}'(P) = P^b$:
\begin{equation}
\frac{DF}{Dt} = G' \cdot {\cal R}(P_{\rm mh}) \cdot {\cal P}'(P) \cdot {\cal F}(F)
\end{equation}
where, measuring $P$ in ${\rm GPa}$, 
\begin{eqnarray}
G' &=& \left( 36 \pi \right)^{1/3} \chi^{1/3} u_{\rm CJ} P_0^{-b} \\
   &\approx& 56\ {\rm ms}^{-1}\,{\rm GPa}^{-b} \left( \frac{\chi}{10^9\,{\rm mm}^{-3}}\right)^{1/3} \times \\
           &\ & \ \ \ \ \ \ \      \times     \left( \frac{u_{\rm CJ}}{10\,{\rm \mu m}\,{\rm \mu s}^{-1}}  \right)
             \left( \frac{P_{\rm CJ}}{30\,{\rm GPa}}  \right)^{-b}
\end{eqnarray}
Collecting this together, what we have then is
\begin{multline}
\frac{DF}{Dt} = {\color{black} G}\,\Phi^{1/3}\!\!\left(\frac{\log{P_{\rm mh}}-\log{\color{black}P_\mu}}{\color{black}\sigma} \right) \times\\
\times \left(\frac{P}{\color{black}P_0}\right)^{\color{black}b} \left[-\log(1-F)\right]^{2/3}(1-F)
\end{multline}
or, as it is used in practice, with $P_0$ is absorbed into $G'$, and highlighting the free parameters in red,
\begin{multline}
\frac{DF}{Dt} = {\color{red} G'}\,\Phi^{1/3}\!\!\left(\frac{\log{P_{\rm mh}}-\log{\color{red}P_\mu}}{\color{red}\sigma} \right) \times\\
\times P^{\color{red}b} \left[-\log(1-F)\right]^{2/3}(1-F)
\end{multline}

What we have described, in a nutshell, is the Salinas reactive flow rate model, which we developed
in mid-2016 and tested in the ALE hydrocode {\tt ares} at LLNL in 2016-2017. It has four free parameters
-- $G', P_\mu, \sigma, b$. (But note that in practice
in all work to date we have fixed $b=2$, and only varied the three remaining parameters.)
 As it turns out,
on close examination, this model is quite similar to the SURF model proposed by Menikoff and Shaw (2010)\cite{SURFa}.

\section{Major Influences}
Souers et al (2004), in studying corner-turning in the TATB-based IHE LX-17,
 noted \cite{Souers_etal_2004}, ``Our conclusion that
modeling [of] failure requires a separate package is new in the reactive flow field and is
not accepted by everyone.''
Here, we describe established models that inspired us to develop {\em Salinas}.

\subsection{Ignition \& Growth}

\begin{figure}
\includegraphics[width=7.0cm]{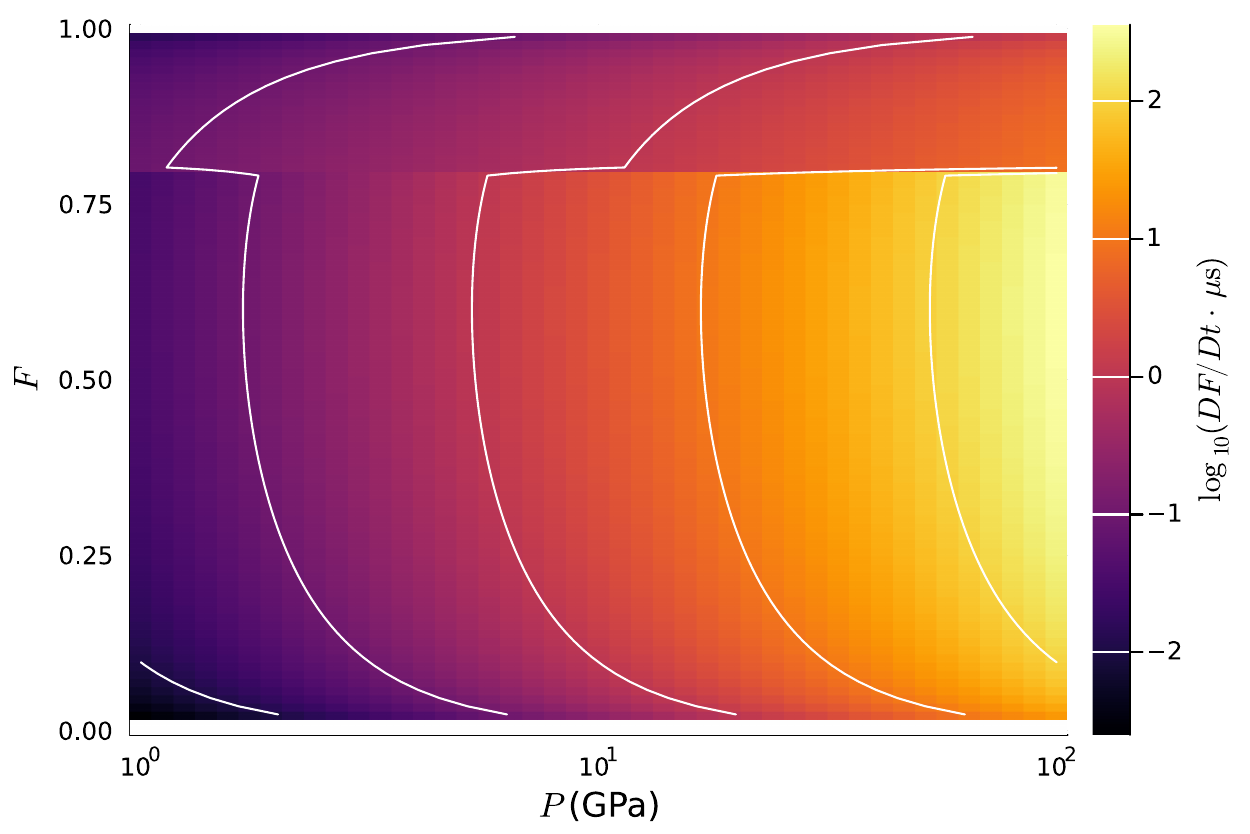}
\caption{Ignition \& Growth rate for LX-17 (growth and completion terms).}
\label{fig:IGrate} 
\end{figure}

Ignition \& Growth (I\&G) \cite{IG, Tarver_2005} is the workhorse reactive-flow model of LLNL --- indeed the original
such model --- with a long pedigree.

The main reaction rate law for I\&G divides the reaction state space up into three different
regions can be written a few different ways, for example \cite{Tarver_2005},
\begin{alignat*}{3}
\frac{DF}{Dt} = \\
{}& I(1-F)^b \left( \frac{\rho}{\rho_0} -1 -a\right)^x \ \ &&0 < F < F_{\rm i} \\ 
 +\, &{\color{red} G_1}(1-F)^{\color{red} c} F^{\color{red}d} P^{\color{red}y}    &&0 < F < {\color{red}F_1} \\ 
  +\, &{\color{red}G_2}(1-F)^{\color{red}e} F^{\color{red}g} P^{\color{red}z}    &&{\color{red}F_2} < F < 1 %
\label{eq:IG}
\end{alignat*}
(Key rate law parameters for growth and completion are highlighted in red.) 
While important, the ignition term (the first term) only covers a small volume of the full $(F,P,\rho)$ state space, since $F_{\rm i}$ is typically rather small ({\em e.g.} $0.02$). The rate law outside of this -- the growth and completion terms -- is functionally of the form
\begin{equation}
\frac{DF}{Dt} = \sum_i G_i \cdot {\cal F}_i(F) \cdot {\cal P}_i(P).
\end{equation}
where the sum runs over a partition of $F$.

In principle, the combined growth and completion terms have ten free parameters, although in practice
not all of these are typically taken advantage of. For example, Tarver provides 
$F_1 = F_2$ and $c=e=g$ for hockey-puck experiments with the TATB-based IHE LX-17 \cite{Tarver_2005}.
One could not argue however that the parameter space is anything less than at least seven-dimensional. This is a strength in that it affords the modeler flexibility to fit a very broad range of
experiments. On the other hand it remains to be seen if it can do so for the same parameter settings, which is what one would
ideally like.

\subsection{Tarantula 2011}
\begin{figure}
\includegraphics[width=7.0cm]{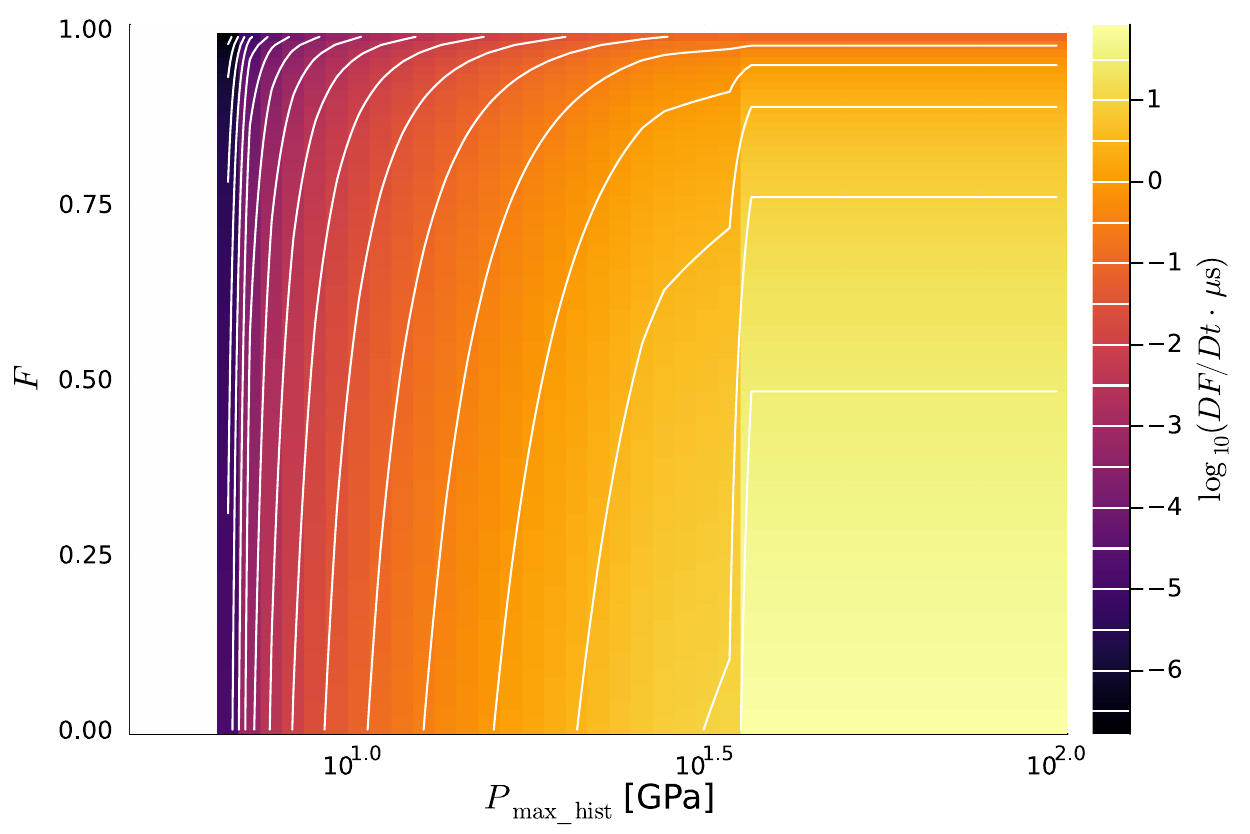}
\caption{Rate in a $(P_{\rm mh},F)$ slice of $(P,P_{\rm mh},F)$ state space for
Tarantula 2011, for $P=P_{\rm mh}$.}
\label{fig:Tarantularate} 
\end{figure}

The Salinas model was heavily influenced by JWL++ Tarantula 2011 [JT11], which is both fast and seemed to fit corner-turning with relative ease. Rate in JT11 is a function of the three-dimensional state space $(P_{\rm mh},P,F)$. Outside of desensitization
(which is not of concern here), the JT11 model rate law may be written
\begin{alignat*}{4}
\frac{DF}{Dt} = \\
&0,& &&  &P& \!< \!P_0 \\ 
&{\color{red}G_1} \left( P_{\rm mh} - {\color{red}P_{o1}}\right)^{\color{red}b_1} (1-F)^{\color{red}c_1},&\ \ \   && {\color{red}P_0}\! \le &P& \!<\! P_1 \\ 
&{\color{red}G_2} \left( P_{\rm mh} - {\color{red}P_{o2}}\right)^{\color{red}b_2} (1-F)^{\color{red}c_2},& \ \ \   && {\color{red}P_1}\! \le &P& \!<\! P_2 \\
&{\color{red}G_3} \left( P_{\rm mh} - {\color{red}P_{o3}}\right)^{\color{red}b_3} (1-F)^{\color{red}c_3},&\ \ \   && {\color{red}P_2}\! \le &P& \\
\label{eq:JT11}
\end{alignat*}
In principle, this rate law appears to have 15 free parameters. In practice, $b_3=0$, $c_1=c_2=1$, $P_{o1}=P_{o2}=P_{o3}=P_0$.
 In the end,
one could argue that the free parameter count is as low as 8 $(G_1,G_3,b_1,b_2,c_3,P_0,P_1,P_2)$. (Note that $G_1$ and $G_2$ are not independent since they appear to be chosen to enforce continuity of the rate law
on the way up when $P=P_{\rm mh}$.)

\begin{figure}
\includegraphics[width=7.0cm]{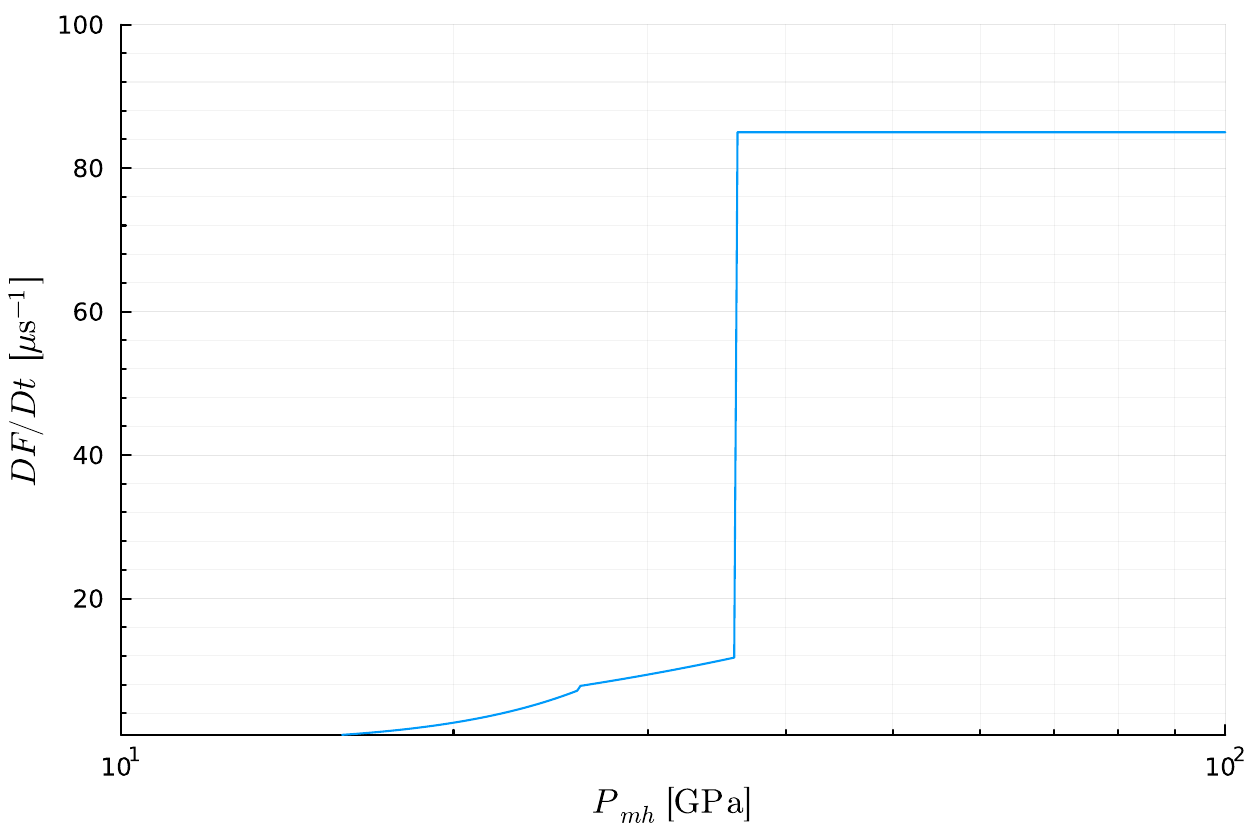}
\caption{Rate in Tarantula 2011, for $F=0$ and $P=P_{\rm mh}$.}
\label{fig:Tarantula1D} 
\end{figure}

\begin{figure}
\includegraphics[width=7.0cm]{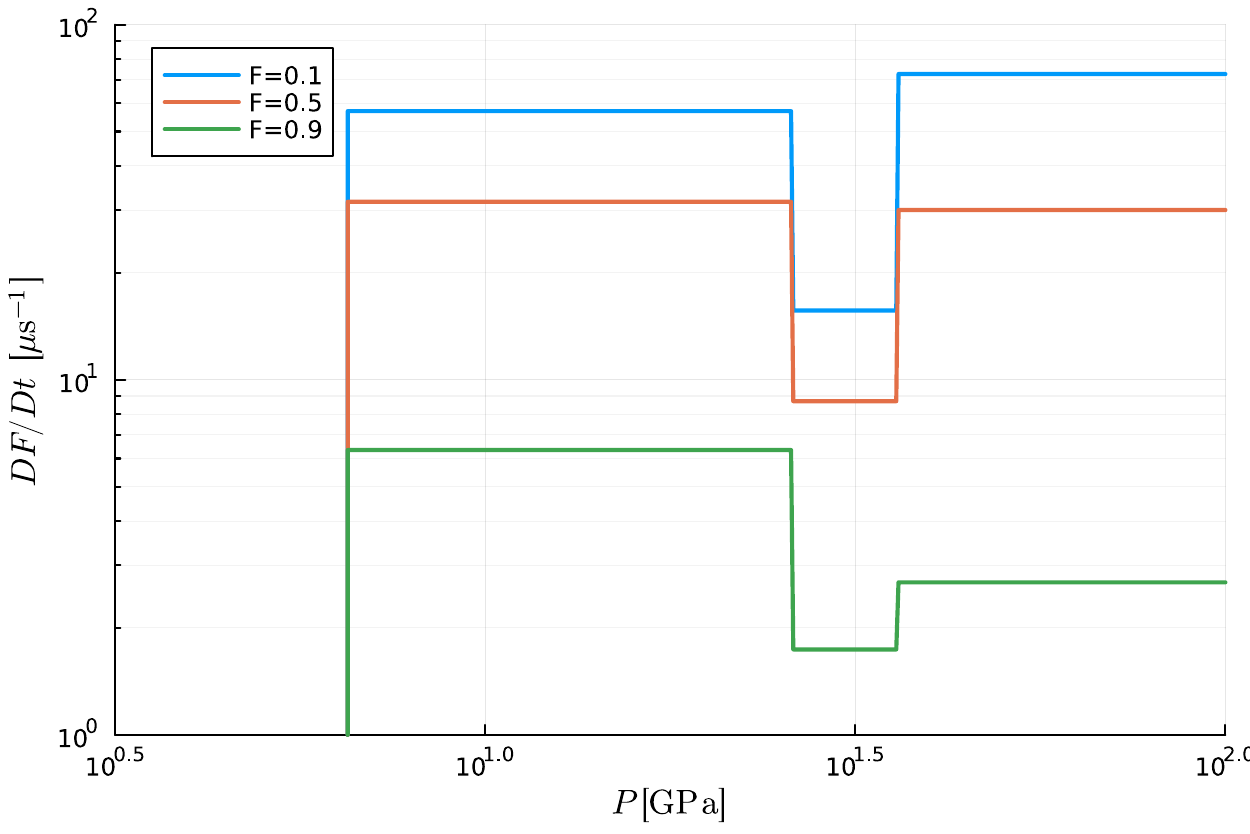}
\caption{Rate in Tarantula 2011, for $P_{\rm mh} = 50\,{\rm GPa}$.}
\label{fig:Tarantula1Dx50} 
\end{figure}

Of greater concern is the aforementioned discontinuity (see figs.~\ref{fig:Tarantula1D},\ref{fig:Tarantula1Dx50}).
note that, given actual parameter values in
use \cite{Tarantula2011}, the rate is strongly discontinuous, with a large step-function jump in rate not only at $P_2$
but at $P_0$ and $P_1$ as well, behind the (potentially) initiating shock.
Experience with CFD suggests that numerical methods are not always compatible with abrupt step-like changes in rate laws.

\begin{figure}
\includegraphics[width=7.0cm]{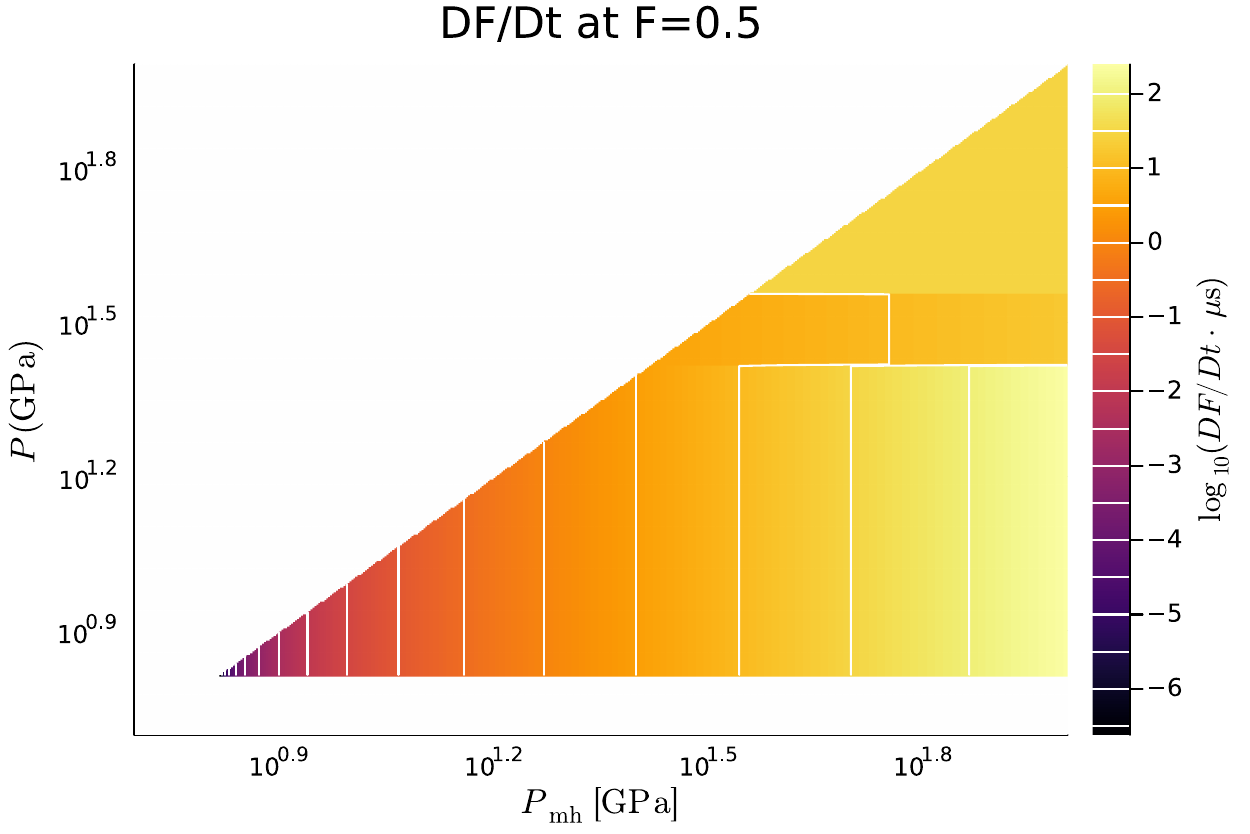}
\caption{Rate in Tarantula 2011, for $F=0.5$.}
\label{fig:TarantulaTriangle} 
\end{figure}

Moreover, the rate is not always monotonic with respect to $P$ when $P < P_{\rm mh}$. This can be seen by visualizing the rate
law in a slice of the full $P_{\rm mh},P,F$ state space at a fixed $F$. Since, by definition, $P\le P_{\rm mh}$, the rate
law only occupies a half-space of the $P_{\rm mh}, P$ plane at any fixed $F$, which appears as a triangular region on a plot;
see fig.~(\ref{fig:TarantulaTriangle}) for $F=0.5$.

This lack of monotonicity may seen quite clearly if we imagine that the initiating shock has already passed,
and we show the rate as a function of $P$ for a range of different $F$. Let us suppose a fully-developed
detonation wave with a von Neumann spike of
about $50\,{\rm GPa}$ (note that we are not claiming the actual von Neumann spike for PBX~9502 is $50\,{\rm GPa}$;
the precise value does not matter much for this exercise, so long as it is well above the CJ pressure).

One can see that not only is the rate non-monotonic with respect to $P$ --- meaning that, in the reaction zone if
not into the Taylor wave, the reaction rate will actually go {\em up} as the pressure goes down --- but that
in fact, for $F > 0.5$, the rate is actually highest in the initiation region ($P_0<P<P_1$) rather than
in the detonation region ($P>P_2$).

\begin{figure}
\includegraphics[width=7.0cm]{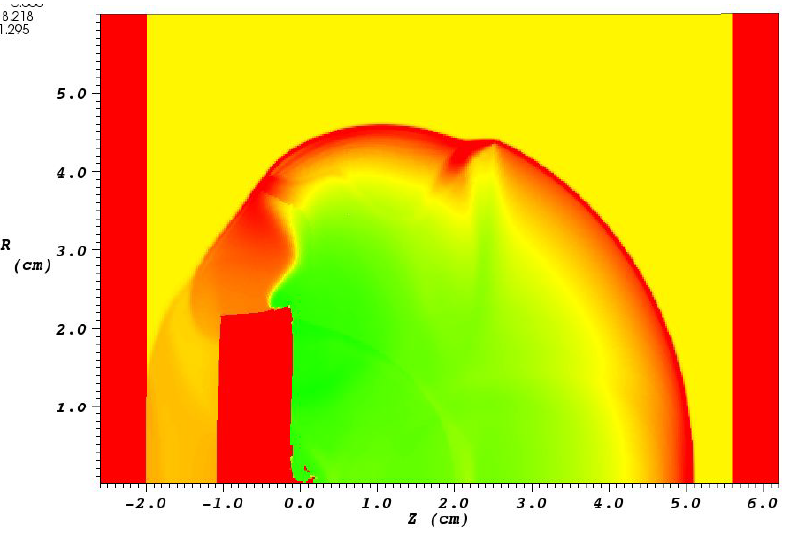}
\caption{Figure 8a from Souers, Haylett, \& Vitello (2011)\cite{Tarantula2011}, showing possible instability 
in burn in expanding detonation wave near $z=2.5\,{\rm cm}$, $R=4.0\,{\rm cm}$.}
\label{fig:wavy} 
\end{figure}

This may be the explanation for persistent instabilities that we encountered in attempting to run
JWL++ Tarantula 2011. One such instance we encountered was a rate stick with light confinement, which led to
alternating bands of partial burn and complete burn as a function of distance along the rate stick. We also encountered
instability in expanding (curved) detonation waves, which were also characterized by regions of incomplete burn.
This may be the explanation for certain ``wavy'' features seen in the founding literature \cite{Tarantula2011} 
for the model; see fig.~({\ref{fig:wavy}}).

Despite this, it is interesting to note the success in capturing corner-turning with a model that makes use
of $P_{\rm mh}$ as key phenomenological state quantity that strongly influences burn subsequent to the initiating
shock. It is also worth noting that, while in general the rate law is a complex function of $F,P,P_{\rm mh}$,
it can be written simply as
\begin{equation}
\frac{DF}{Dt} = G \cdot {\cal R}(P_{\rm mh}) \cdot {\cal F}(F)
\end{equation}
on the way up to the peak shock pressure, {\em i.e.} when $P=P_{\rm mh}$. This inspired us to consider generalizing
this to the ansatz
\begin{equation}
\frac{DF}{Dt} = G \cdot {\cal R}(P_{\rm mh}) \cdot {\cal P}(P) \cdot {\cal F}(F)
\end{equation}
that was the basis for Salinas.

\subsection{Cheetah + pqplf2}
The other model at LLNL that we drew inspiration from was a model used by the energetics materials group there to
study ECOT, an experiment developed at LANL\cite{ECOT}.
This model used the CHEETAH thermochemical code\cite{CHEETAH}, making use of a piecewise-linear rate law invoked by 
a call to the CHEETAH routine {\em pqplf2} \cite{PQPLF2}.

\begin{figure}
\includegraphics[width=7.0cm]{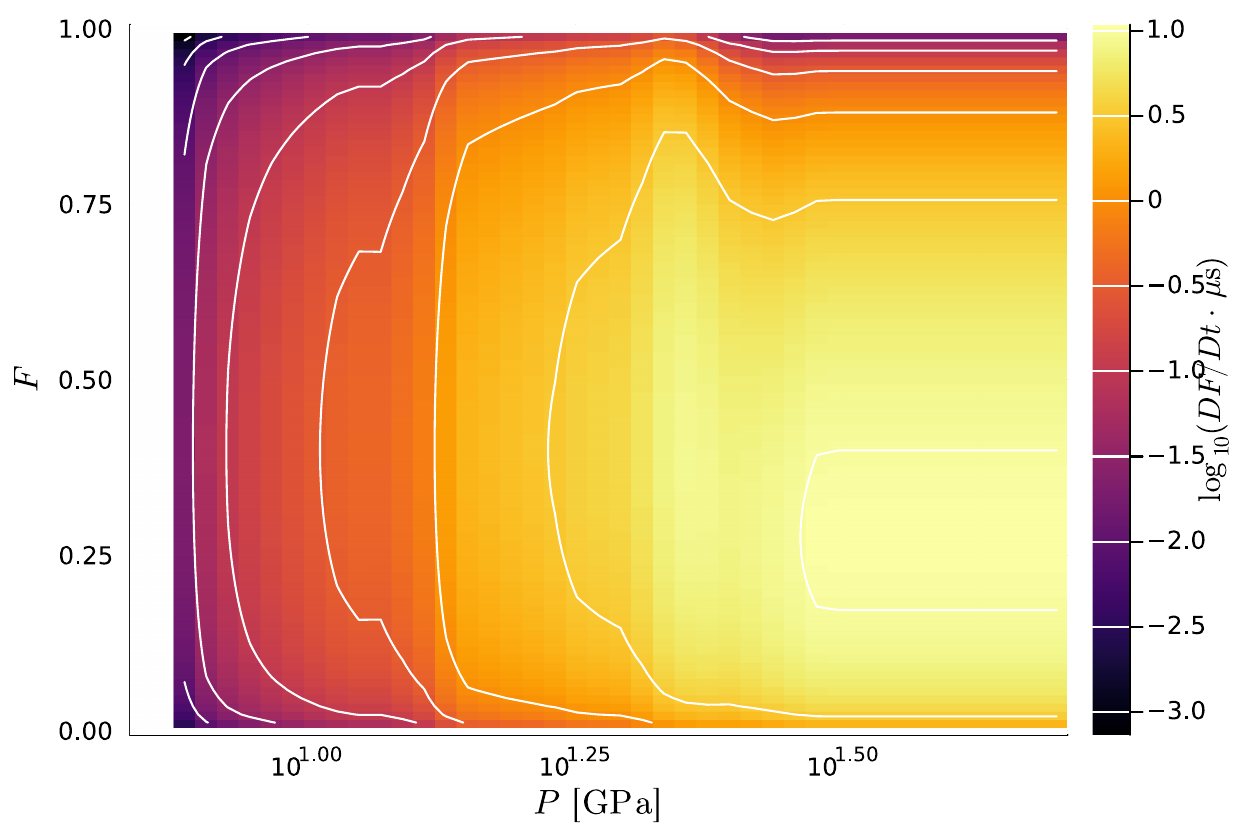}
\caption{Example rate in $(P,F)$ state space for a {\em pqplf2} rate law.}
\label{fig:pqplf2rate} 
\end{figure}

The rate is written
\begin{equation}
\frac{DF}{Dt} = {\color{red}f}(P) \cdot (1-F)^{1+{\color{red}C}(P)} F^{\color{red}b}
\end{equation}
but now $f$ and $C$ are piecewise-linear functions of $P$. These curves are specified
by a table of approximately 20--25 indicial values ${P_i}$ at which there are 20--25 tabulated
values ${f_i}$ and 20--25 tabulated values ${C_i}$. On its face, this would indicate somewhere
between 61 (20+20+20+1) and 76 (25+25+25+1) free parameters. In actual fact, the spacing of
the $P_i$ is mostly uniform in the bulk, and most $C_i$ below about $23\,{\rm GPa}$ are taken to 
be $0$, so the number of effective free parameters is substantially lower, as low as the low 30's or so.
Still, this is not a small number.

This poses a problem when combined with the fact that, for us at least with the computational resources
available to us, it took approximately 8 hours for a single run to compare with ECOT (for example). This means
effectively that it is not possible to perform any sort of systematic exploration of the parameter space to
find an optimal fit to a single experiment, let alone to a suite of different experiments.

This inspired us to ask if we could develop a model that was, on the one hand, as fast as JWL++ Tarantula 2011,
but was stable and had even fewer free parameters.

\begin{figure}
\includegraphics[width=7.0cm]{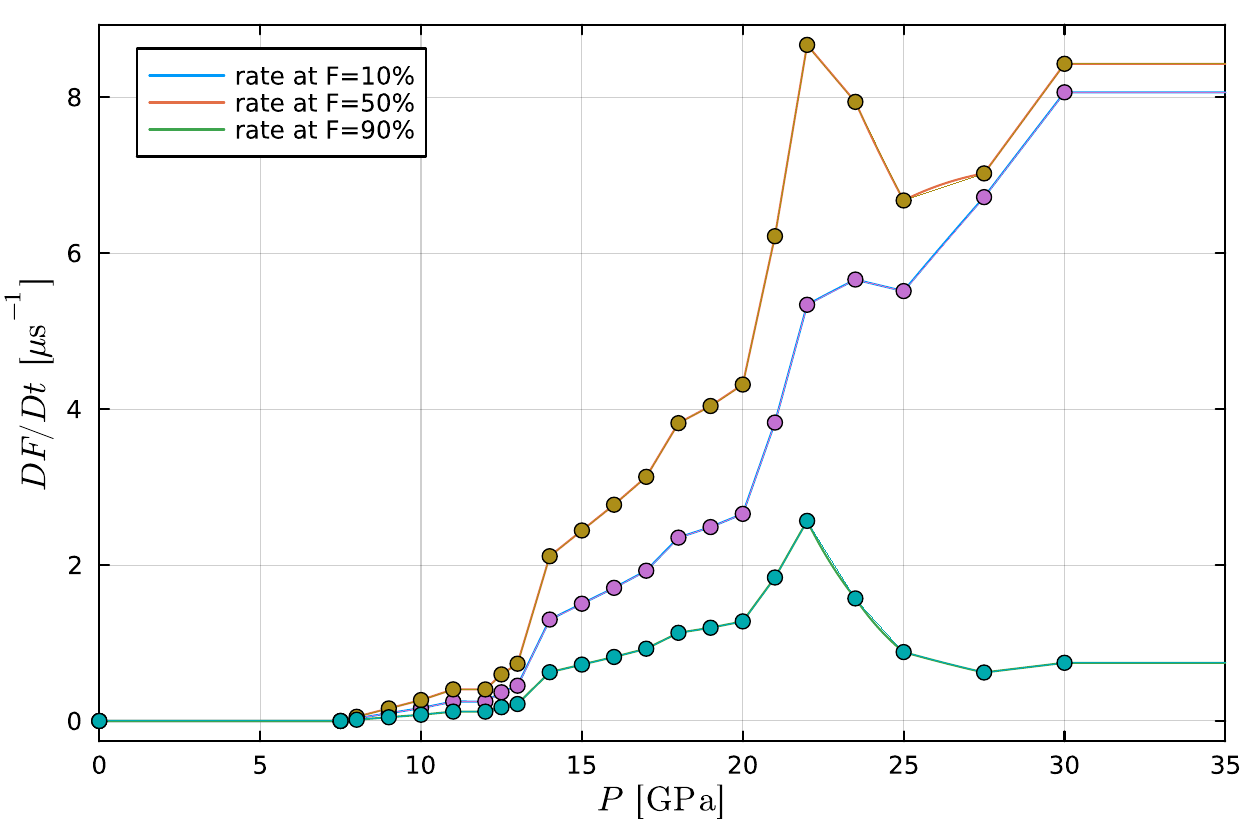}
\caption{Example rate versus $P$ for $F=0.1,\,0.5,0.9$ for a {\em pqplf2} rate law (Model 1.29/2017).}
\label{fig:pqplf2rate2} 
\end{figure}

%

\section{Results}
\begin{figure}
\includegraphics[width=7.0cm]{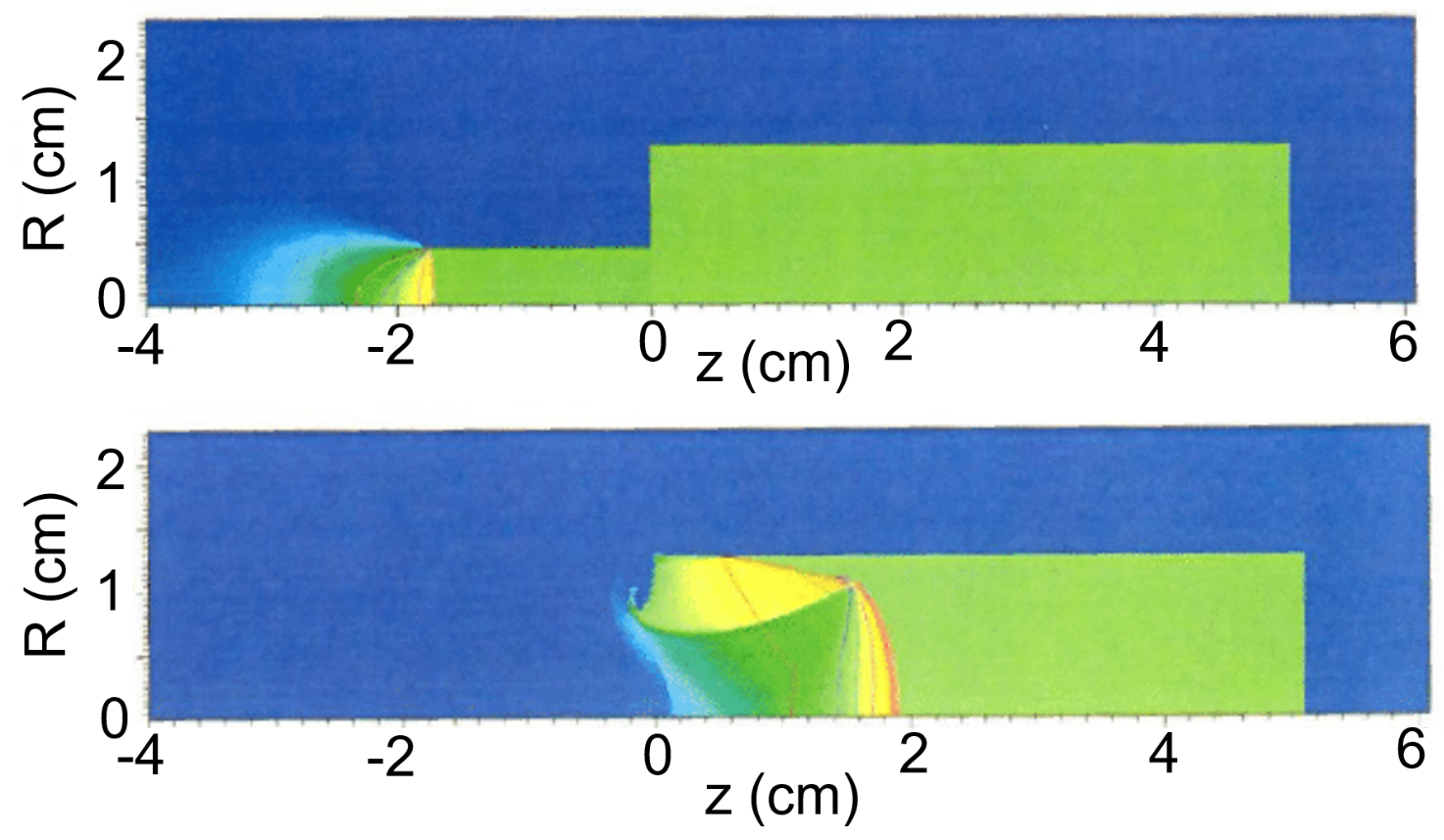}
\caption{Intermediate stages of burn for a simulation of ECOT using Salinas.}
\label{fig:ECOTsim} 
\end{figure}

\begin{figure}
\includegraphics[width=7.0cm]{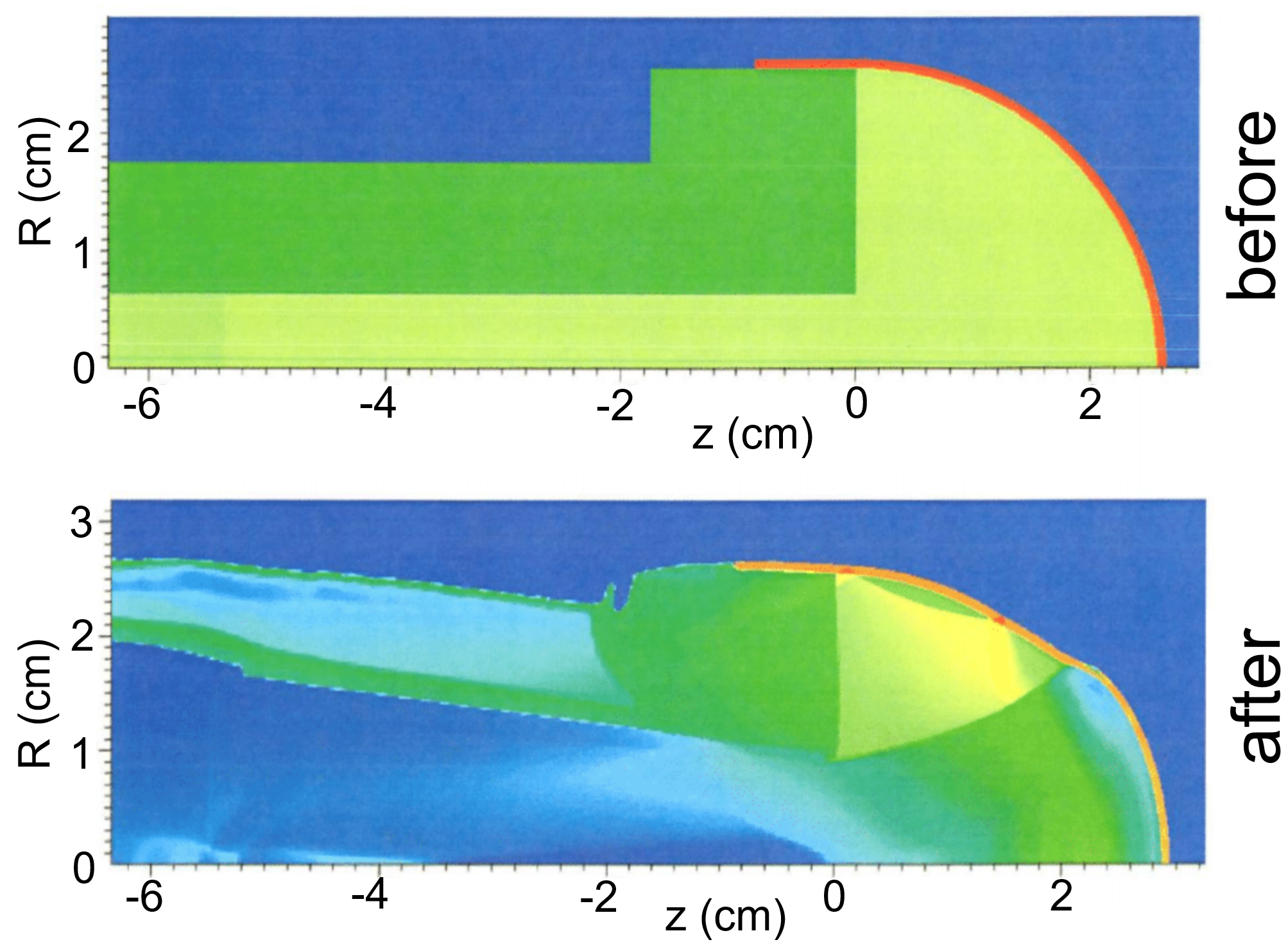}
\caption{Intermediate stages of burn for a simulation of SAX using Salinas.}
\label{fig:SAXsim} 
\end{figure}

We were able to run Salinas in a variety of experiments. The full suite of experiments
(ECOT, SCOT, DAX, SAX) is described in Williams (2020) \cite{Will_2020}. Simulation of 
ECOT may be seen in fig.~(\ref{fig:ECOTsim}). Simulation of SAX is seen
in fig.~(\ref{fig:SAXsim}).

Unfortunately, we are no longer in posession of our data showing the fits of Salinas to
the experimental data for ECOT, SCOT, DAX or SAX. However, we welcome any opportunities
that may be available to incorporate Salinas into other reactive-flow hydrodynamic codes
so that we may perform further work on the model and demonstrate its fidelity in 
a range of experiments including, but not limited to, corner-turning experiments
for PBX~9502.

\section{Acknowledgments}
The author thanks Edward L. Lee and K. Thomas Lorenz for their encouragement.

\bibliographystyle{Det_Symp}

\bibliography{IDS_paper_}

\end{document}